\documentclass[12pt]{article}

\usepackage[pdftex,  
    plainpages=false,
    bookmarks         = true,
    bookmarksnumbered = true,
    pdfpagemode       = None,
    pdfstartview      = FitH,
    pdfpagelayout     = SinglePage,
    colorlinks        = false,
    linkcolor         = blue,
    pagecolor         = black,
    urlcolor          = blue, 
    pdfborder         = {0 0 0}
    ]{hyperref}%

\usepackage{graphicx}
\usepackage{cite}
\usepackage{color}
\usepackage{colortbl}
\usepackage{xspace}
\usepackage{rotating}

\begin{document}


\newcommand{\comment}[1]{\par\noindent {\em\small [#1]}}
\renewcommand{\comment}[1]{}


\newcommand{\mysubsection}[1]{\subsection[#1]{\boldmath #1}}

\newcommand{\unit}[1]{\ensuremath{\rm\,#1}}
\newcommand{\keV}{\unit{keV}}
\newcommand{\keVc}{\unit{keV/{\it c}}}
\newcommand{\keVcc}{\unit{keV/{\it c}^2}}
\newcommand{\MeV}{\unit{MeV}}
\newcommand{\MeVc}{\unit{MeV\!/\!{\it c}}}
\newcommand{\MeVcc}{\unit{MeV\!/\!{\it c}^2}}
\newcommand{\GeV}{\unit{GeV}}
\newcommand{\GeVc}{\unit{GeV\!/\!{\it c}}}
\newcommand{\GeVcc}{\unit{GeV\!/\!{\it c}^2}}
\newcommand{\TeV}{\unit{TeV}}
\newcommand{\invfb}{\unit{fb^{-1}}}
\newcommand{\cm}{\unit{cm}}
\newcommand{\mm}{\unit{mm}}
\newcommand{\mb}{\unit{mb}}
\newcommand{\ps}{\unit{ps}}
\newcommand{\invps}{\unit{ps^{-1}}}
\newcommand{\fs}{\unit{fs}}
\newcommand{\microns}{\unit{\mu m}}
\newcommand{\mrad}{\unit{mrad}}
\newcommand{\rad}{\unit{rad}}
\newcommand{\Hz}{\unit{Hz}}
\newcommand{\kHz}{\unit{kHz}}
\newcommand{\MHz}{\unit{MHz}}



\newcommand{\CP}{\ensuremath{\rm CP}}
\newcommand{\bra}[1]{\ensuremath{\langle #1|}}
\newcommand{\ket}[1]{\ensuremath{|#1\rangle}}
\newcommand{\braket}[2]{\ensuremath{\langle #1|#2\rangle}}

\newcommand{\IP}{\ensuremath{\rm IP}}
\newcommand{\signif}[1]{\ensuremath{\rm #1/\sigma_{#1}}}
\newcommand{\signifm}[1]{\ensuremath{#1/\sigma_{#1}}}
\newcommand{\pT}{\ensuremath{p_{\rm T}}}
\newcommand{\pTmin}{\ensuremath{\pT^{\rm min}}}
\newcommand{\ET}{\ensuremath{E_{\rm T}}}
\newcommand{\DLL}[2]{\ensuremath{\rm \Delta  \ln{\cal L}_{\particle{#1}\particle{#2}}}}
\newcommand{\BR}[1]{\ensuremath{\rm BR(#1)}}
\newcommand{\BRvis}[1]{\ensuremath{\rm BR_{\rm vis}(#1)}}

\newcommand{\dm}{\ensuremath{\Delta m}}
\newcommand{\dms}{\ensuremath{\Delta m_{\rm s}}}
\newcommand{\dmd}{\ensuremath{\Delta m_{\rm d}}}
\newcommand{\DG}{\ensuremath{\Delta\Gamma}}
\newcommand{\DGs}{\ensuremath{\Delta\Gamma_{\rm s}}}
\newcommand{\DGd}{\ensuremath{\Delta\Gamma_{\rm d}}}
\newcommand{\Gs}{\ensuremath{\Gamma_{\rm s}}}
\newcommand{\Gd}{\ensuremath{\Gamma_{\rm d}}}

\newcommand{\MBq}{\ensuremath{M_{\Bq}}}
\newcommand{\DGq}{\ensuremath{\Delta\Gamma_{\rm q}}}
\newcommand{\Gq}{\ensuremath{\Gamma_{\rm q}}}
\newcommand{\dmq}{\ensuremath{\Delta m_{\rm q}}}
\newcommand{\GL}{\ensuremath{\Gamma_{\rm L}}}
\newcommand{\GH}{\ensuremath{\Gamma_{\rm H}}}

\newcommand{\DGsGs}{\ensuremath{\Delta\Gamma_{\rm s}/\Gamma_{\rm s}}}
\newcommand{\Dm}{\mbox{$\Delta m $}}
\newcommand{\ACP}{\ensuremath{{\cal A}^{\rm CP}}}
\newcommand{\Adir}{\ensuremath{{\cal A}^{\rm dir}}}
\newcommand{\Amix}{\ensuremath{{\cal A}^{\rm mix}}}
\newcommand{\ADelta}{\ensuremath{{\cal A}^\Delta}}
\newcommand{\phid}{\ensuremath{\phi_{\rm d}}}
\newcommand{\sinphid}{\ensuremath{\sin\!\phid}}
\newcommand{\phis}{\ensuremath{\phi_{\rm s}}}
\newcommand{\betas}{\ensuremath{\beta_{\rm s}}}
\newcommand{\sbetas}{\ensuremath{\sigma(\beta_{\rm s})}}
\newcommand{\stbetas}{\ensuremath{\sigma(2\beta_{\rm s})}}
\newcommand{\sinphis}{\ensuremath{\sin\!\phis}}

\newcommand{\vOmega}{\ensuremath{\vec{\Omega}}}
\newcommand{\vkappa}{\ensuremath{\vec{\kappa}}}
\newcommand{\vlambda}{\ensuremath{\vec{\lambda}}}

\newcommand{\Af}{\ensuremath{A_{f}}}
\newcommand{\Afbar}{\ensuremath{\overline{A}_{\overline{f}}}}
\newcommand{\fbar}{\ensuremath{\BAR{ f}}}

\newcommand{\calH}{\ensuremath{{\cal H}}}
\newcommand{\tauBs}{\ensuremath{\tau_\Bs}}
\newcommand{\tauL}{\ensuremath{\tau_{\rm L}}}
\newcommand{\tauH}{\ensuremath{\tau_{\rm H}}}

\newcommand{\effs}{{\ensuremath{\varepsilon^{\rm s}_\theta }}}
\newcommand{\effb}{{\ensuremath{\varepsilon^{\rm b}_\theta}}}

\newcommand{\Vud}{\ensuremath{V_{\rm ud}}}
\newcommand{\Vus}{\ensuremath{V_{\rm us}}}
\newcommand{\Vub}{\ensuremath{V_{\rm ub}}}
\newcommand{\Vcd}{\ensuremath{V_{\rm cd}}}
\newcommand{\Vcs}{\ensuremath{V_{\rm cs}}}
\newcommand{\Vcb}{\ensuremath{V_{\rm cb}}}
\newcommand{\Vtd}{\ensuremath{V_{\rm td}}}
\newcommand{\Vts}{\ensuremath{V_{\rm ts}}}
\newcommand{\Vtb}{\ensuremath{V_{\rm tb}}}
\newcommand{\VCKM}{\ensuremath{V_{\rm CKM}}}

\newcommand{\Vjr}{\ensuremath{V_{ jr}}}
\newcommand{\Vjb}{\ensuremath{V_{j {\rm b}}}}

\newcommand{\edet}{{\ensuremath{\varepsilon_{\rm det}}}}
\newcommand{\erec}{{\ensuremath{\varepsilon_{\rm rec/det}}}}
\newcommand{\esel}{{\ensuremath{\varepsilon_{\rm sel/rec}}}}
\newcommand{\etrg}{{\ensuremath{\varepsilon_{\rm trg/sel}}}}
\newcommand{\etot}{{\ensuremath{\varepsilon_{\rm tot}}}}

\newcommand{\mistag}{\ensuremath{\omega}}
\newcommand{\etag}{{\ensuremath{\varepsilon_{\rm tag}}}}
\newcommand{\effeff}{\ensuremath{\varepsilon_{\rm eff}}}
\newcommand{\efftag}{{\ensuremath{\etag(1-2\omega)^2}}}
\newcommand{\effD}{{\ensuremath{\etag D^2}}}

\newcommand{\etagprompt}{{\ensuremath{\varepsilon_{\rm tag}^{\rm prompt}}}}
\newcommand{\etagLL}{{\ensuremath{\varepsilon_{\rm tag}^{\rm LL}}}}

\newcommand{\tage}{\particle{e}}
\newcommand{\tagmu}{\particle{\mu}}
\newcommand{\tagKopp}{\particle{K_{opp}}}
\newcommand{\tagKsame}{\particle{K_{same}}}
\newcommand{\tagPsame}{\particle{\pi_{same}}}
\newcommand{\tagQvtx}{\ensuremath{Q_{\rm vtx}}}
\newcommand {\mch}      {\multicolumn {4} {|c|}}


\newcommand{\BAR}[1]{\overline{#1}}

\newcommand{\particle}[1]{{\ensuremath{\rm #1}}}

\newcommand{\pp}{\particle{pp}}
\newcommand{\ppbar}{\particle{p\BAR{p}}}

\newcommand{\cc}{\particle{c\BAR{c}}}
\renewcommand{\b}{\particle{b}}
\newcommand{\bbar}{\particle{\BAR{b}}}
\newcommand{\bb}{\particle{b\BAR{b}}}

\newcommand{\B}{\particle{B}}
\newcommand{\Bd}{\particle{B^0}}
\newcommand{\Bs}{\particle{B^0_s}}
\newcommand{\Bds}{\particle{B^0_{(s)}}}
\newcommand{\Bu}{\particle{B^+}}
\newcommand{\Bc}{\particle{B^+_c}}
\newcommand{\Lb}{\particle{\Lambda_b}}

\newcommand{\Bbar}{\particle{\BAR{B}}}
\newcommand{\Bdbar}{\particle{\BAR{B}{^0}}}
\newcommand{\Bsbar}{\particle{\BAR{B}{^0_s}}}
\newcommand{\Bdsbar}{\particle{\BAR{B}{^0}_{(s)}}}
\newcommand{\Bubar}{\particle{B^-}}
\newcommand{\Bcbar}{\particle{B^-_c}}
\newcommand{\Lbbar}{\particle{\BAR{\Lambda}_b}}

\newcommand{\Bqq}{\particle{B_q}}
\newcommand{\Bqqbar}{\particle{\BAR{B_q}}}

\newcommand{\BL}{\particle{B_L}}
\newcommand{\BH}{\particle{B_H}}
\newcommand{\BLH}{\particle{B_{L,H}}}

\newcommand{\Ds}{\particle{D_s}}
\newcommand{\Dsm}{\particle{D_s^-}}
\newcommand{\KKpim}{\particle{K^+K^-\pi^-}}
\newcommand{\KK}{\particle{KK}}
\newcommand{\KpKm}{\particle{K^+K^-}}
\newcommand{\Dsp}{\particle{D_s^+}}
\newcommand{\Dsmp}{\particle{D_s^{\mp}}}

\newcommand{\Bq}{\particle{B_q}}
\newcommand{\Bqbar}{\particle{\BAR{B}_{q}}}

\newcommand{\Dz}{\particle{D^0}}
\newcommand{\Dzbar}{\particle{\BAR{D}{^0}}}
\newcommand{\DzCP}{\particle{D^0_{CP}}}
\newcommand{\Dstar}{\particle{D^{*-}}}
\newcommand{\Dstarz}{\particle{D^{0*}}}
\newcommand{\DstInc}{\particle{D^{(*)-}}}

\newcommand{\Jpsi}{\particle{J\!/\!\psi}}
\newcommand{\Jmm}{\particle{\Jpsi(\mu\mu)}}
\newcommand{\Jee}{\particle{\Jpsi(ee)}}
\newcommand{\ee}{\particle{e^+e^-}}
\newcommand{\JpsiX}{\particle{\Jpsi X}}

\newcommand{\KS}{\particle{K^0_S}}  
\newcommand{\Kst}{\particle{K^{*0}}}  

\newcommand{\Kplus}{\particle{K^+}}
\newcommand{\Kminus}{\particle{K^-}}
\newcommand{\Kpm}{\particle{K^\pm}}

\newcommand{\pip}{\particle{\pi^+}}
\newcommand{\pim}{\particle{\pi^-}}
\newcommand{\piz}{\particle{\pi^0}}

\newcommand{\Jphi}{\particle{J\!/\!\psi\phi}}
\newcommand{\JKK}{\particle{J\!/\!\psi\KpKm}}

\newcommand{\BsBs}{\Bs--\Bsbar}


\newcommand{\decay}[2]{\particle{#1\!\to #2}}

\newcommand{\bccsbar}{\decay{\BAR{b}}{\BAR{c}c\BAR{s}}}

\newcommand{\KSpipi}{\decay{K^0_S}{\pi^+\pi^-}}  
\newcommand{\Jpsiee}{\decay{\Jpsi}{e^+e^-}}  
\newcommand{\Jpsimm}{\decay{\Jpsi}{\mu\mu}}  
\newcommand{\Jpsill}{\decay{\Jpsi}{\ell^+\ell^-}}  

\newcommand{\Kpi}{\particle{K^+\pi^-}}
\newcommand{\pippim}{\particle{\pi^+\pi^-}}

\newcommand{\DsKKpi}{\decay{\Dsm}{K^+K^-\pi^-}}  
\newcommand{\phiKK}{\decay{\phi}{K^+K^-}}
\newcommand{\KstKpi}{\decay{\Kst}{\Kpi}}

\newcommand{\Bdpipi}{\decay{\Bd}{\pi^+\pi^-}}            
\newcommand{\BdKpi}{\decay{\Bd}{K^+\pi^-}}               
\newcommand{\BspiK}{\decay{\Bs}{\pi^+K^-}}               
\newcommand{\BsKK}{\decay{\Bs}{K^+K^-}}                  
\newcommand{\Bhh}{\decay{\Bds}{h^+h^-}}                  
\newcommand{\BsDspi}{\decay{\Bs}{\Dsm\pi^+}}             
\newcommand{\BsDsK}{\decay{\Bs}{\Dsmp K^{\pm}}}          
\newcommand{\BsDsmKp}{\decay{\Bs}{\Dsm K^+}}             
\newcommand{\BsDspKm}{\decay{\Bs}{\Dsp K^-}}             
\newcommand{\BsDsh}{\decay{\Bs}{\Dsm h^+}}               
\newcommand{\BdJmmKS}{\decay{\Bd}{\Jmm\KS(\pi\pi)}}              
\newcommand{\BdJeeKS}{\decay{\Bd}{\Jee\KS}}              
\newcommand{\BdJKS}{\decay{\Bd}{\Jpsi\KS}}               
\newcommand{\BdJmmKst}{\decay{\Bd}{\Jmm\Kst(\particle{K}\pi)}}            
\newcommand{\BdJeeKst}{\decay{\Bd}{\Jee\Kst}}            
\newcommand{\BdJKst}{\decay{\Bd}{\Jpsi\Kst}}             
\newcommand{\BuJmmK}{\decay{\Bu}{\Jmm K^+}}              
\newcommand{\BuJeeK}{\decay{\Bu}{\Jee K^+}}              
\newcommand{\BuJK}{\decay{\Bu}{\Jpsi K^+}}               

\newcommand{\BsDsmu}{\decay{\Bs}{\Dsm\mu^+\nu}}          

\newcommand{\BuJX}{\decay{\Bu}{\Jpsi X}}  
\newcommand{\BdJX}{\decay{\Bd}{\Jpsi X}}  
\newcommand{\BsJX}{\decay{\Bs}{\Jpsi X}}  
\newcommand{\LbJX}{\decay{\Lb}{\Jpsi X}}  
\newcommand{\BuJmmX}{\decay{\Bu}{\Jmm X}}  
\newcommand{\BdJmmX}{\decay{\Bd}{\Jmm X}}  
\newcommand{\BsJmmX}{\decay{\Bs}{\Jmm X}}  
\newcommand{\LbJmmX}{\decay{\Lb}{\Jmm X}}  
\newcommand{\bJX}{\decay{b}{\Jpsi X}}  

\newcommand{\BsJmmphi}{\decay{\Bs}{\Jmm\phi(\particle{KK})}}            
\newcommand{\BsJeephi}{\decay{\Bs}{\Jee\phi}}            
\newcommand{\BsJeephiKK}{\decay{\Bs}{\Jee\phi(\particle{KK})}}            
\newcommand{\BsJphi}{\decay{\Bs}{\Jpsi\phi}}             
\newcommand{\BsJKK}{\decay{\Bs}{\Jpsi\KpKm}}             

\newcommand{\Bsmm}{\decay{\Bs}{\mu^+\mu^-}}              
\newcommand{\BdmmKst}{\decay{\Bd}{\mu^+\mu^-\Kst}}       
\newcommand{\BdeeKst}{\decay{\Bd}{e^+e^-\Kst}}           
\newcommand{\BdllKst}{\decay{\Bd}{\ell^+\ell^-\Kst}}     
\newcommand{\BdphiKS}{\decay{\Bd}{\phi\KS}}              
\newcommand{\Bsphiphi}{\decay{\Bs}{\phi\phi}}            
\newcommand{\BsKstKst}{\decay{\Bs}{\Kst\Kst}}           
\newcommand{\BdDstpi}{\decay{\Bd}{D^{*-}\pi^+}}          
\newcommand{\BdDstpiincl}{\decay{\Bd}{D^{*-}(incl)\pi^+}}
\newcommand{\BdDKst}{\decay{\Bd}{\Dz\Kst}}               
\newcommand{\BdDbarKst}{\decay{\Bd}{\BAR{D}{^0}\Kst}}    
\newcommand{\BdDCPKst}{\decay{\Bd}{\DzCP\Kst}}           
\newcommand{\BdDbarKpiKst}{\decay{\Bd}{\Dzbar(K\pi)\Kst}}
\newcommand{\BdDbarKKKst}{\decay{\Bd}{\Dzbar(KK)\Kst}}   
\newcommand{\BdDbarpipiKst}{\decay{\Bd}{\Dzbar(\pi\pi)\Kst}}   
\newcommand{\BdDCPKKKst}{\decay{\Bd}{\DzCP(KK)\Kst}}   
\newcommand{\BdKstrho}{\decay{\Bd}{K^{*0}\rho/\omega}}   
\newcommand{\Bsetacphi}{\decay{\Bs}{\eta_c\phi}}         
\newcommand{\BsetacKKphi}{\decay{\Bs}{\eta_c(4K)\phi}}   
\newcommand{\BsetacpiKphi}{\decay{\Bs}{\eta_c(2pi2K)\phi}}  
\newcommand{\Bsetacpipiphi}{\decay{\Bs}{\eta_c(4pi)\phi}}
\newcommand{\BsJmmeta}{\decay{\Bs}{\Jmm\eta}}            
\newcommand{\BsJeeeta}{\decay{\Bs}{\Jee\eta}}            
\newcommand{\BsJeta}{\decay{\Bs}{\Jpsi\eta}}             
\newcommand{\BdKstgam}{\decay{\Bd}{K^{*0}\gamma}}        
\newcommand{\BdKstpin}{\decay{\Bd}{K^{*0}\pi^0}}         
\newcommand{\Bsphigam}{\decay{\Bs}{\phi\gamma}}          
\newcommand{\Bsphipin}{\decay{\Bs}{\phi\pi^0}}           
\newcommand{\Bdpipipi}{\decay{\Bd}{\pi^+\pi^-\pi^0}}     
\newcommand{\Bdrhopi}{\decay{\Bd}{\rho\pi}}              

\newcommand{\BcJmmpi}{\decay{\Bc}{\Jmm\pi^+}}            
\newcommand{\BcJpi}{\decay{\Bc}{\Jpsi\pi^+}}             
\newcommand{\Bsmumu}{\decay{\Bs}{\mu^+\mu^-}}            

\newcommand{\BsDsstDsst}{\decay{\Bs}{\Ds^{(*)-}\Ds^{(*)+}}}  
\newcommand{\BsDsDs}{\decay{\Bs}{\Dsm\Dsp}}             
\newcommand{\BdDstmu}{\decay{\Bd}{\Dstar \mu^+\nu_\mu}}   %

\newcommand{\BudsJX}{\decay{\rm B_{\rm u,d,s}}{\Jpsi \rm X}}


\newcommand{\SM}{Standard Model}             

\newcommand{\fsig}{\ensuremath{f_{\rm sig}}}


\newcommand{\mbs}{m_{\Bs}}   

\newcommand{\Bl}{\particle{B^0_L}}
\newcommand{\Bh}{\particle{B^0_H}}
\newcommand{\Gl}{\Gamma_L}
\newcommand{\Gh}{\Gamma_H}
\newcommand{\delone}{\delta_1}
\newcommand{\deltwo}{\delta_2}
\newcommand{\delperp}{\delta_{\perp}}
\newcommand{\delpar}{\delta_{\|}}
\newcommand{\delzero}{\delta_0}

\def\thetatwo{\theta_2}
\def\apar{A_{\|}(t)}
\def\aperp{A_{\perp}(t)}
\def\azero{A_{0}(t)}
\def\aparO{A_{\|}(0)}
\def\aperpO{A_{\perp}(0)}
\def\azeroO{A_{0}(0)}
\newcommand{\Rt}{R_\perp}
\newcommand{\Rp}{R_\|}
\newcommand{\Ro}{R_0}
\newcommand{\dGtwo}{\frac{\DG}{2}}
\renewcommand{\d}{\textrm{d}}


\newcommand{\thetatr}{\ensuremath{\theta}}
\newcommand{\phitr}{\ensuremath{\varphi}}
\newcommand{\psitr}{\ensuremath{\psi}}
\newcommand{\thetaone}{\ensuremath{\psi}}

\newcommand{\phijphi}{\ensuremath{\Phi_{\Jphi}}}
\newcommand{\Phiphiphi}{\ensuremath{\Phi_{\Bsphiphi}}}
\newcommand{\myphis}{\ensuremath{\Phi}}

\newcommand{\phiM}{\ensuremath{\Phi_{\rm M}}}
\newcommand{\phiD}{\ensuremath{\Phi_{\rm D}}}
\newcommand{\phiCP}{\ensuremath{\Phi_{\rm CP}}}
\newcommand{\phiMNP}{\ensuremath{\Phi^{\rm NP}_{\rm M}}}
\newcommand{\phiDNP}{\ensuremath{\Phi^{\rm NP}_{\rm D}}}
\newcommand{\phiMG}{\ensuremath{\Phi_{\rm M/\Gamma}}}
\newcommand{\phiMGSM}{\ensuremath{\Phi_{\rm M/\Gamma}^{\rm SM}}}

\newcommand{\phipen}{\ensuremath{\Phi_{\rm penguin}}}
\newcommand{\phibox}{\ensuremath{\Phi_{\rm box}}}

\newcommand{\phiNP}{\ensuremath{\Phi^{\rm NP}}}
\newcommand{\phiSM}{\ensuremath{\Phi^{\rm SM}}}
\newcommand{\phisDelta}{\ensuremath{\phi^\Delta_{\rm s}}}
\newcommand{\phitot}{\ensuremath{\Phi_{\rm tot}}}

\newcommand{\diffX}{ \frac{d^{4} \Gamma}{dt d\Omega} }
\newcommand{\diffXbar}{ \frac{d^{4} \bar \Gamma}{dt d\Omega} }

\newcommand{\Ps}{\ensuremath{{\cal P}_{s}}}
\newcommand{\Psbar}{\ensuremath{\bar{{\cal P}_{s}}}}

%
%
\setcounter{footnote}{0}
\renewcommand{\thefootnote}{\alph{footnote}}

\begin{titlepage}

\pagenumbering{arabic}
\vspace*{-1.5cm}
\begin{tabular*}{15.cm}{lc@{\extracolsep{\fill}}r}
\end{tabular*}
\vspace*{1.cm}
\begin{center}
\Large
{\bf \boldmath
\Large {\bf Determination of  $2\betas$  in \decay{\Bs}{\Jpsi \KpKm} Decays in the Presence of a \KpKm\ S-Wave Contribution  }
 \\
\vspace*{1.cm}
 }
\normalsize

{ Yuehong~Xie,\footnote{Yuehong.Xie@cern.ch} 
  Peter~Clarke,\footnote{peter.clarke@ed.ac.uk} 
  Greig~Cowan\footnote{g.cowan@ed.ac.uk} and 
  Franz~Muheim\footnote{f.muheim@ed.ac.uk} \\

\vspace*{0.5cm}
{\it  School of Physics and Astronomy, University of Edinburgh, \\
       Mayfield Road, Edinburgh, EH9 3JZ, U.K. }\\

 }
\end{center}
\vspace{\fill}
\begin{abstract}
\noindent

\normalsize

We present  the complete differential decay rates for the process  \decay{\Bs}{\Jpsi \KpKm}
including S-wave and P-wave angular momentum states for the \KpKm\ meson pair. 
We examine the effect of  an S-wave component on the determination of
the CP violating phase   $2\betas$. Data from the B-factories indicate that  an S-wave component of 
about $10\%$ may be expected  in the $\phi(1020)$ resonance region.
We find that if this contribution is ignored in the analysis it could cause a bias in the measured value of $2\betas$ towards zero of
the order of $10\%$.
When including the \KpKm\ S-wave component we observe an increase in the statistical error on $2\betas$ by less than $15\%$.
We also point out the possibility of  measuring the sign of  $\cos2\betas$ by using the interference between the \KpKm\ S-wave and P-wave amplitudes 
to resolve the strong phase ambiguity. 
We conclude that the S-wave component can be properly taken into account in the analysis.

\end{abstract}
\vspace*{2.cm}
\begin{center}
\Large {\bf \boldmath\huge {\bf }}
\end{center}
\vspace{\fill}
\end{titlepage}

\setcounter{page}{1}
\cleardoublepage
\hrule

\tableofcontents

\vspace*{1cm}

\hrule

\vspace*{1cm}


\linewidth=\columnwidth

%

\setcounter{footnote}{0}
\renewcommand{\thefootnote}{\arabic{footnote}}

\section{Introduction}
\label{sec:intro}
The decay \BsJphi\ is a golden channel for the measurement of the \Bs\ mixing phase  $-2\betas$  which is
a very sensitive probe of new physics.
It has been extensively studied~\cite{Buchalla:2008jp,Ball:2006xx,ref:lhcb-2007-101,Ciuchini:2006dw,Lenz:2006hd,Ligeti:2006pm, CDF, D0}.
In the decay \BsJphi, followed by a two-body decay \decay{\phi(1020)}{\KpKm},  
the \KpKm\ meson pair is in an orbital P-wave amplitude.
However, in the vicinity of the  \particle{\phi (1020)} mass,
the \KpKm\ system can have contributions from other partial waves.
The same comment holds for the \KpKm\ system in the decay channels 
 \decay{\Bd}{\KpKm\KS} and  \decay{\Dz}{\KpKm \piz}.
The BaBar experiment  showed that in these decays
the S-wave and P-wave contributions dominate
in the mass range above  threshold up to  $1.1\GeVcc$~\cite{BabarB2KKKs, BabarD2KKpi}. 
In both cases there is a dominant resonant  \particle{\phi (1020)}  contribution. In addition an S-wave  \particle{f_0(980)} and a non-resonant contribution
are found to be necessary to describe the data.
These results motivated us to investigate the effects of a possible
S-wave contribution to \decay{\Bs}{\Jpsi \KpKm} in the  \particle{\phi (1020)}  mass region.

In  the decay \decay{\Bs}{\Jpsi \KpKm} the \KpKm\  system 
can only arise from a ${\rm s\BAR{s}}$ quark pair
while in \decay{\Bd}{\KpKm \KS} and  \decay{\Dz}{\KpKm \piz}
it can have contributions from both ${\rm s\BAR{s}}$  and  ${\rm d\BAR{d}}$.
This makes it difficult to give a quantitative estimate
for the S-wave component.
In reference~\cite{swave-sheldon} the S-wave \KpKm\ contribution under the
 \particle{\phi (1020)} peak is estimated to be $5-10\%$ for decay modes
in which the $\Kplus \Kminus$  arises from an ${\rm s\bar{s}}$ quark pair. 
In this study we consider an S-wave of similar magnitude
and assess its impact on the determination of  the weak mixing phase $-2\betas$.

\section{Time-dependent angular distributions in the decay \decay{\Bs}{\Jpsi\KpKm} 
including  S-wave contributions}\label{secDist}
\setcounter{equation}{0}

We consider P- and S-wave amplitudes in the decay \decay{\Bs}{\Jpsi\KpKm} 
 where the invariant mass of  the \KpKm\ meson pair is in the \particle{\phi(1020)} mass region
and the \Jpsi\ meson decays into a $\particle{\mu^+\mu^-}$ pair. 
The S-wave contribution can be non-resonant or due to the  
\particle{f_0(980)} resonance\footnote{The mass dependence of the \particle{f_0(980)} 
is distorted as the central value of the resonance is below threshold.}.
We denote  decay amplitudes for the   \decay{\Bs}{\Jpsi \KpKm}  
by  $ {\bf \boldmath{A}}=(A_0,A_{||},A_{\perp},A_S)$.
 Here $A_0$, $A_{||}$ and $A_{\perp}$ are the three P-wave amplitudes consistent
 with the \KpKm\ system decaying via the \particle{\phi(1020)} resonance.
 $A_S$ is the  amplitude  for a possible S-wave  contribution in the \KpKm\ system.
 The amplitudes for the conjugate decay  \decay{\Bsbar}{\Jpsi \KpKm}  
 are denoted by  $ {\bf \boldmath{\bar A}}=(\bar A_0, \bar A_{||},\bar A_{\perp},\bar A_S)$,
which, in the absence of direct CP violation, are related to $ {\bf \boldmath{A}}$ by  
$A_0 = \bar A_0$, $A_{||}= \bar A_{||}$, $A_{\perp}= -\bar A_{\perp}$ and $A_S= - \bar A_S$.
Note that $A_0$ and $A_{||}$ are CP-even whereas  $A_{\perp}$ and $A_S$ are CP-odd.
The amplitudes  $ (A_0,A_{||},A_{\perp})$ and 
the amplitude $A_S $ may have different dependences
on the mass  $m_{\KpKm}$ of the \KpKm\ system. 
However, in sufficiently small bins of  $m_{\KpKm}$, such as the narrow mass region around  
the \particle{\phi (1020)} resonance,  the dependences of the amplitudes  on $m_{\KpKm}$ 
can be neglected.

We define the total P-wave strength, $ A_P^2 \equiv |A_0|^2+|A_{||}|^2+|A_{\perp}|^2$,
the longitudinal and perpendicular polarisation fractions relative to the P-wave strength
$R_{||} \equiv |A_{||}|^2 / A_P^2$,
and $R_{\perp} \equiv |A_{\perp}|^2 / A_P^2$,
and the S-wave fraction,
$R_{S} \equiv |A_{S}|^2 / (A_P^2+|A_S|^2)$.
The phases of these decay amplitudes are defined by
$A_j = |A_j| e^{i\delta_j}$, where $j=0, ||, \perp, S$.
As only the relative strong phase differences can be measured we adopt the convention $\delta_0 = 0$.

An angular analysis is required to disentangle the different CP eigenstates
on a statistical basis.
The angular observables are  denoted as the helicity angles $\Omega = (\theta_l, \theta_K, \varphi)$.
Here $\theta_l$ is the angle between the $\mu^+$ momentum and the direction opposite
to the \Bs\ momentum in the \Jpsi\ rest frame;
$\theta_K$ is the angle between the K$^{+}$ momentum and the direction opposite
to the \Bs\ momentum in the rest frame of the \KpKm\ system;
$\varphi$ is the angle between the decay planes  of the \decay{ \Jpsi}{\mu^+\mu^-} and the  \KpKm\ pair,
when going from  the positive kaon to the positive lepton with a 
rotation around the opposite direction of the \Bs\ momentum in the \Jpsi\ rest frame.

The differential decay rate for a \Bs\ meson produced at time $t=0$ decaying as
\decay{\Bs}{\Jpsi \KpKm} at proper time $t$ is given by

\begin{equation}\label{Eqbsrate}
  \frac{\d^{4} \Gamma(\BsJKK) }{\d t \; \d\cos\thetatr \;  \d\cos\thetaone \; \d \varphi} 
\propto  \sum^{10}_{k=1} h_k(t) f_k( \Omega) \,,
\end{equation}
whereas the differential decay rate for an initial \Bsbar\ meson is given by

\begin{equation}\label{EqbsbarRate}
  \frac{\d^{4} \Gamma(\Bsbar\to\JKK) }{\d t \; \d\cos\thetatr \; \d\cos\thetaone \; \d \varphi} 
\propto \sum^{10}_{k=1} \bar{h_k}(t) f_k (\Omega ) \,.
\end{equation}
Each of the $h_k(t)$, $\bar{h_k}(t)$ and $f_k(\Omega)$ for $k=1-10$ are 
defined in Table~\ref{tabAngDef}. In total there are four amplitude-squared terms for the three polarisations
of the P-waves and the S-wave component plus six interference terms.

\begin{table}[h]
\begin{center}
  \begin{tabular}{|c|c|c|c|}
  \hline
    {\small $k$}&
    {\small $h_k(t)$}&
    {\small $\bar{h_k}(t)$}&
    {\small $f_k(\theta_l,\theta_K, \varphi)$}\tabularnewline
    \hline
    {\small 1}&
    {\small ${|A_0(t)|^2}$}&
    {\small ${|\bar{A}_0(t)|^2}$}&
    {\small $4 \sin^2{\theta_l} \cos ^2 {\theta_K} $}\tabularnewline
    \hline
    {\small 2}&
    {\small ${|A_{||}(t)|^2}$}&
    {\small ${|\bar{A}_{||}(t)|^2}$}&
    {\small $(1+\cos^2 {\theta_l}) \sin^2 {\theta_K} -
              \sin^2 {\theta_l} \sin^2 {\theta_K} \cos {2\varphi} $}\tabularnewline
    \hline
    {\small 3}&
    {\small ${|A_{\perp}(t)|^2}$}&
    {\small ${|\bar{A}_{\perp}(t)|^2}$}&
    {\small $(1+\cos^2 {\theta_l}) \sin^2 {\theta_K} +
              \sin^2 {\theta_l} \sin^2 {\theta_K} \cos {2\varphi} $}\tabularnewline
    \hline
    {\small 4}&
    {\small $\Im\{{A^{*}_{||}(t)A_{\perp}(t)}\}$}&
    {\small $\Im\{{\bar{A}^{*}_{||}(t)\bar{A}_{\perp}(t)}\}$}&
    {\small $ 2 \sin^2 {\theta_l} \sin^2 {\theta_K} \sin {2 \varphi}$}\tabularnewline
    \hline
    {\small 5}&
    {\small $\Re\{{A^{*}_0(t)A_{||}(t)}\}$}&
    {\small $\Re\{{\bar{A}^{*}_0(t)\bar{A}_{||}(t)}\}$}&
    {\small $-\sqrt{2} \sin {2 \theta_l} \sin {2 \theta_K} \cos {\varphi}$}\tabularnewline
    \hline
    {\small 6}&
    {\small $\Im\{{A^{*}_0(t)A_{\perp}(t)}\}$}&
    {\small $\Im\{{\bar{A}^{*}_0(t)\bar{A}_{\perp}(t)}\}$}&
    {\small $\sqrt{2} \sin {2\theta_l} \sin {2 \theta_K} \sin {\varphi}$}\tabularnewline

    \hline
    {\small 7}&
    {\small $ |A_S(t)|^2$}&
    {\small $|\bar A_S(t)|^2$}&
    {\small $\frac{4}{3}\sin^2 {\theta_l} $}\tabularnewline

    \hline
    {\small 8}&
    {\small $\Re\{{A^{*}_S(t)A_{||}(t)}\}$}&
    {\small $\Re\{{\bar{A}^{*}_S(t)\bar{A}_{||}(t)}\}$}&
    {\small $-\frac{2}{3} \sqrt {6} \sin{2 \theta_l} \sin{ \theta_K} \cos{\varphi}$}\tabularnewline

    \hline
    {\small 9}&
    {\small $\Im\{{A^{*}_S(t)A_{\perp}(t)}\}$}&
    {\small $\Im\{{\bar{A}^{*}_S(t)\bar{A}_{\perp}(t)}\}$}&
    {\small $\frac{2}{3} \sqrt {6}\sin{2 \theta_l} \sin{ \theta_K} \sin{\varphi}$}\tabularnewline

    \hline
    {\small 10}&
    {\small $\Re\{{A^{*}_S(t)A_0(t)}\}$}&
    {\small $\Re\{{\bar{A}^{*}_S(t)\bar{A}_0(t)}\}$}&
    {\small $\frac{8}{3} \sqrt {3}\sin ^2{ \theta_l} \cos { \theta_K} $}\tabularnewline
    \hline

  \end{tabular}
\end{center}
\caption{Definition of the functions $h_k(t)$, $\bar{h_k}(t)$ and  $f_k(\theta_l,\theta_K, \varphi)$ of Eq.~\ref{Eqbsrate} 
and~\ref{EqbsbarRate}.}
\label{tabAngDef}
\end{table}

The  time-dependence of the ten functions $h_k(t)$ 
for an initial $\Bs$ meson state can be written as:

\footnotesize
\begin{eqnarray}
  |A_0(t)|^2 &=&
  |A_0|^2 {\rm e}^{-\Gs t} \Bigl[ \cosh\left(\frac{\DGs t}{2}\right)
   - \cos\myphis\sinh\left(\frac{\DGs t}{2}\right)
   \: + \: \sin\myphis\sin(\dms t) \Bigr] \,, \label{ab1} \\
  |A_{\|}(t)|^2 &=&
  |A_{\|}|^2 {\rm e}^{-\Gs t} \Bigl[ \cosh\left(\frac{\DGs t}{2}\right)
   - \cos\myphis \sinh\left(\frac{\DGs t}{2}\right)
  \: + \: \sin\myphis\sin(\dms t) \Bigr] \,,  \label{ab2} \\
  |A_{\perp}(t)|^2 &=&
  |A_{\perp}|^2 {\rm e}^{-\Gs t} \Bigl[ \cosh\left(\frac{\DGs t}{2}\right)
   + \cos\myphis \sinh\left(\frac{\DGs t}{2}\right)  \: - \: \sin\myphis\sin(\dms t) \Bigr] \,,  \label{ab3} \\
    \Im\{A_{\|}^*(t)A_{\perp}(t)\} &=&
    |A_{\|}||A_{\perp}|  {\rm e}^{-\Gs t}   \Bigl[
    - \cos(\delperp-\delpar)\sin\myphis\sinh\left(\frac{\DGs t}{2}\right) \nonumber \\
    & & \: + \: \sin(\delperp-\delpar)\cos(\dms t) \: - \: \cos(\delperp-\delpar) \cos\myphis \sin(\dms t) \Bigr] \,,  \label{ab4} \\
  \Re\{A_0^*(t) A_{\|}(t)\} &=&
  |A_0||A_{\|}| {\rm e}^{-\Gs t} \cos(\delpar -\delta_0)
  \Bigl[  \cosh\left(\frac{\DGs t}{2}\right) - \cos\myphis \sinh
  \left(\frac{\DGs t}{2}\right) \nonumber \\
  & &  \: + \: \sin\myphis \sin(\dms t) \Bigr]  \,,  \label{ab5} \\
   \Im\{A_{0}^*(t)A_{\perp}(t)\} &=&
   |A_{0}||A_{\perp}|  {\rm e}^{-\Gs t} \Bigl[
    - \cos (\delperp -\delta_0)\sin\myphis\sinh\left(\frac{\DGs t}{2}\right) \nonumber  \label{ab6} \\
    & & \: + \: \sin(\delperp -\delta_0 )\cos(\dms t) \: - \:
   \cos(\delperp -\delta_0)\cos\myphis\sin(\dms t) \Bigr] \,, \\
  |A_{S}(t)|^2 &=&
  |A_{S}|^2 {\rm e}^{-\Gs t} \Bigl[ \cosh\left(\frac{\DGs t}{2}\right)
   + \cos\myphis \sinh\left(\frac{\DGs t}{2}\right)  \: - \: \sin\myphis\sin(\dms t) \Bigr] \,,  \label{ab7} \\
   \Re\{A_{S}^*(t)A_{\|}(t)\} &=&
    |A_{S}||A_{\|}|  {\rm e}^{-\Gs t}   \Bigl[
    - \sin(\delpar -\delta_S)\sin\myphis\sinh\left(\frac{\DGs t}{2}\right) \nonumber \\
    & & \: + \: \cos (\delpar -\delta_S)\cos(\dms t) \: - \: \sin(\delpar -\delta_S) \cos\myphis \sin(\dms t) \Bigr] \,,  \label{ab8} \\
   \Im\{A_{S}^*(t)A_{\perp}(t)\} &=&
  |A_S||A_{\perp}| {\rm e}^{-\Gs t} \sin(\delperp -\delta_S)
  \Bigl[  \cosh\left(\frac{\DGs t}{2}\right) + \cos\myphis \sinh
  \left(\frac{\DGs t}{2}\right) \nonumber \\
  & &  \: - \: \sin\myphis \sin(\dms t) \Bigr]  \,,  \label{ab9} \\
   \Re\{A_{S}^*(t)A_{0}(t)\} &=&
    |A_{S}||A_{0}|  {\rm e}^{-\Gs t}   \Bigl[
    - \sin(\delta_0 -\delta_S)\sin\myphis\sinh\left(\frac{\DGs t}{2}\right) \nonumber \\
    & & \: + \: \cos (\delta_0 -\delta_S)\cos(\dms t) \: - \: \sin(\delta_0 -\delta_S) \cos\myphis \sin(\dms t) \Bigr] \,,  \label{ab10} 
\end{eqnarray}
\normalsize
where $\myphis=-2\betas$, \dms,  \DGs\ and \Gs\ denote the  weak mixing phase, mass difference, decay width 
difference and average decay width of the \Bs-\Bsbar\ system, respectively. 
Here we have assumed that each of the decay amplitudes in $ {\bf \boldmath{A}}$ is dominated by a single 
weak phase,
therefore a common effective $2\betas$ can be used for all CP eigenstates.  
The time evolution functions $\bar{h_k}(t)$ for an initial \Bsbar\ meson can be obtained by reversing the sign
of each term proportional to $\sin(\dms t)$ or $\cos (\dms t)$    in ${h_k}(t)$.

\section{Measuring $2\betas$ in the presence of a \KpKm\ S-wave }\label{secImpact}

In this section we investigate how the measurement of  $2\betas$  
is affected by the presence of a possible \KpKm\ S-wave contribution.
We use Monte Carlo simulated toy data based on the differential decay rate expressions of  Section~\ref{secDist}.
We generate signal decays only and ignore backgrounds underneath the \Bs\ mass peak  as well as
all  detector effects. The inclusion of these effects does not alter the qualitative results of this study.

We assume a tagging efficiency $\epsilon_{tag} =56\%$ and a wrong tag probability $\omega_{tag}=33\%$, 
which correspond approximately to the expected flavour tagging performance
for this channel at the LHCb experiment~\cite{tagging}. 
In Table~\ref{tabPara} we summarize the values of the physical parameters used to
generate the toy data sets. 

We generate 500 data sets for different  scenarios where we vary the values of the S-wave fraction $R_S$ and its 
phase $\delta_S$ and  the weak phase $-2\betas$. 
Each data set contains 30000 signal events corresponding to approximately one quarter of a nominal
LHCb year of  2\invfb .

\begin{table}[h]
\footnotesize
\begin{center}
  \begin{tabular}{|c|c|c|c|c|c|c|c|c|c|c|c|}
  \hline
   & \dms\ & \Gs\ & \DGs\ & $\delta_0$ & $\delta_{\|}$ &  $\delta_{\perp}$ & $R_{\|}$ & $R_{\perp}$ & $R_S$ & $\delta_S$ & $2\betas$  \tabularnewline
  \hline
Input  & $17.8\,\ps^{-1}$ & $0.68\,\ps^{-1}$  & $0.05\,\ps^{-1}$ & 0 & -2.93 & 2.91  & 0.207 & 0.233 & vary & vary & vary \tabularnewline
  \hline
Fit   & fix & fix & fix & fix & float & float & float & float & float$^*$ & float & float \tabularnewline
  \hline
  \end{tabular}
\end{center}
$^*R_S$ is fixed to 0 when the S-wave component is neglected.

\normalsize

\caption{Values of the physical parameters used in the generation of signal decays 
and how these parameters are treated in the fit. }
\label{tabPara}
\end{table}

We perform fits to each data set 
where $2\betas$, $R_S$, $\delta_S$, $R_{||}$, $\delta_{||}$, $R_{\perp}$, $\delta_{\perp}$
are free parameters and all other parameters are kept fixed. 
We also perform fits where the S-wave component is present in the generated toy data, but ignored  in the fit ($R_S$ is set to 0) 
in order to investigate the bias in the determination of $2\betas$.

\begin{figure}[ht]
    \vfill\begin{minipage}{0.5\linewidth}
   \includegraphics[angle=90,width=80mm]{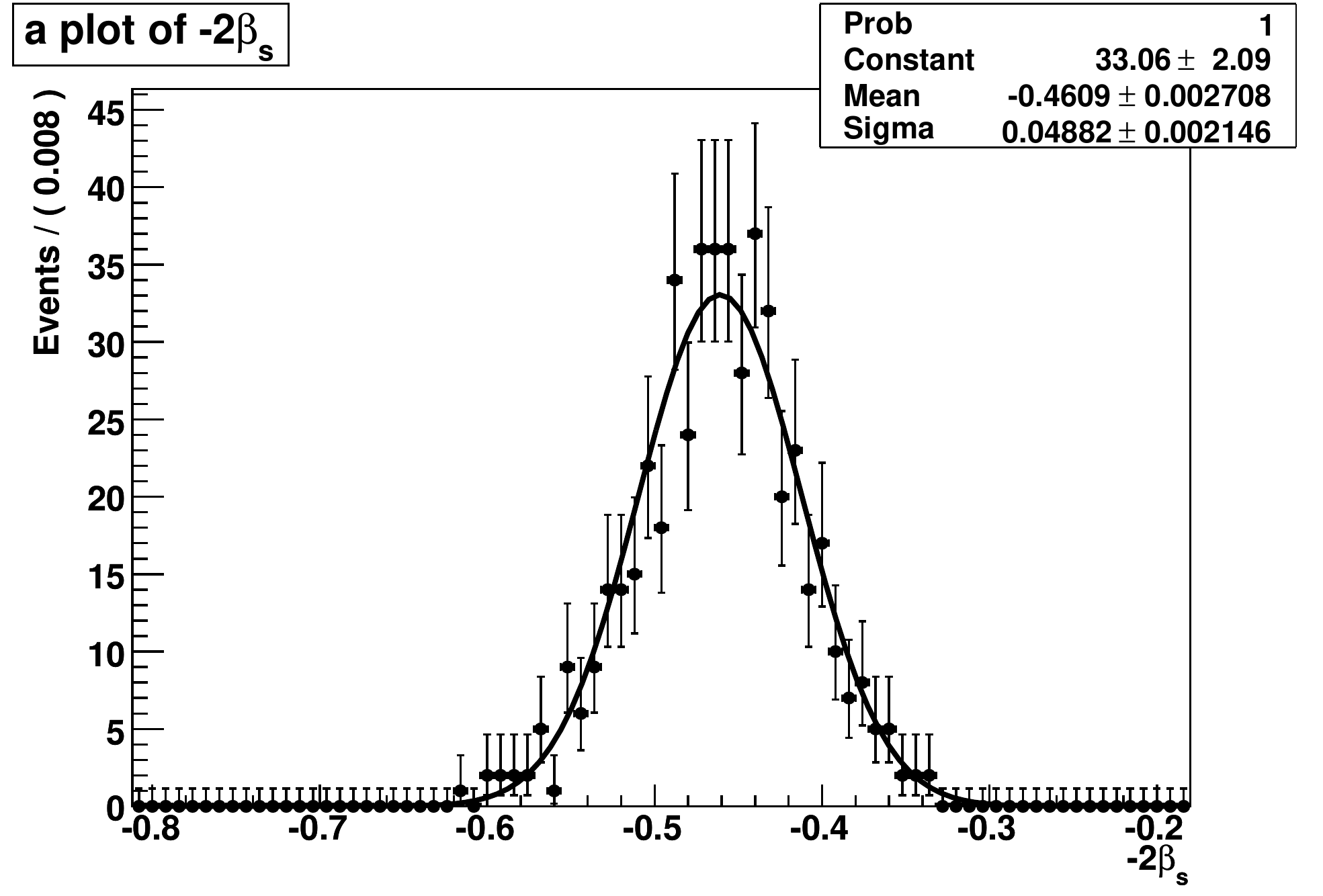}
\end{minipage}
\begin{minipage}{0.5\linewidth}
   \includegraphics[angle=90,width=80mm]{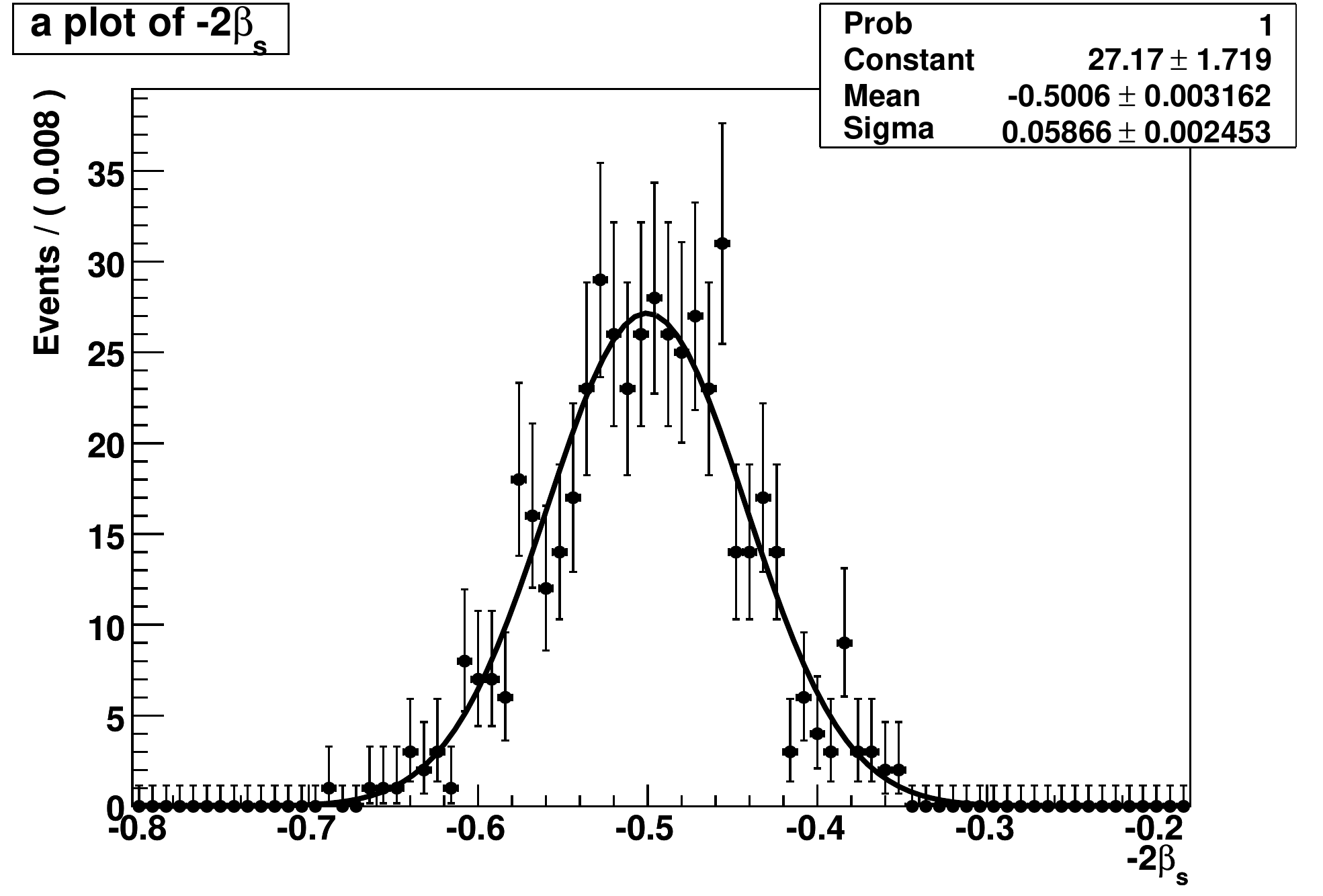}
\end{minipage}
\caption{Distributions of  the fitted values of $-2\beta_s$ for the scenario $R_S=0.1, \delta_S=\pi/2, 2\betas =0.5$. 
The left and right plots are obtained with or without fixing $R_S$ to 0 in fitting the data, respectively. }
\label{phisFit}
\end{figure}

\begin{figure}[ht] 
    \vfill\begin{minipage}{0.5\linewidth}
   \includegraphics[angle=90,width=80mm]{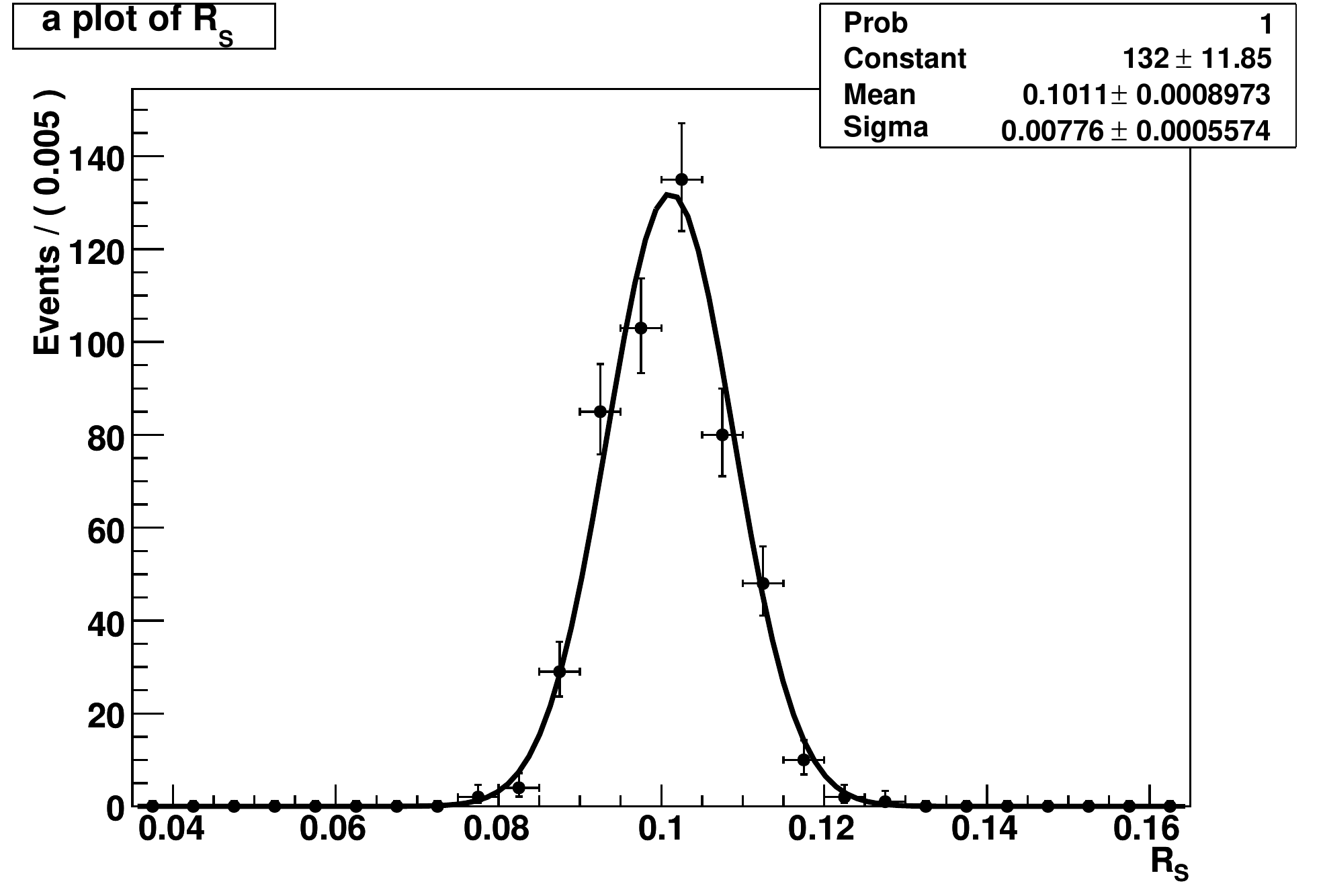}
\end{minipage}
\begin{minipage}{0.5\linewidth}
   \includegraphics[angle=90,width=80mm]{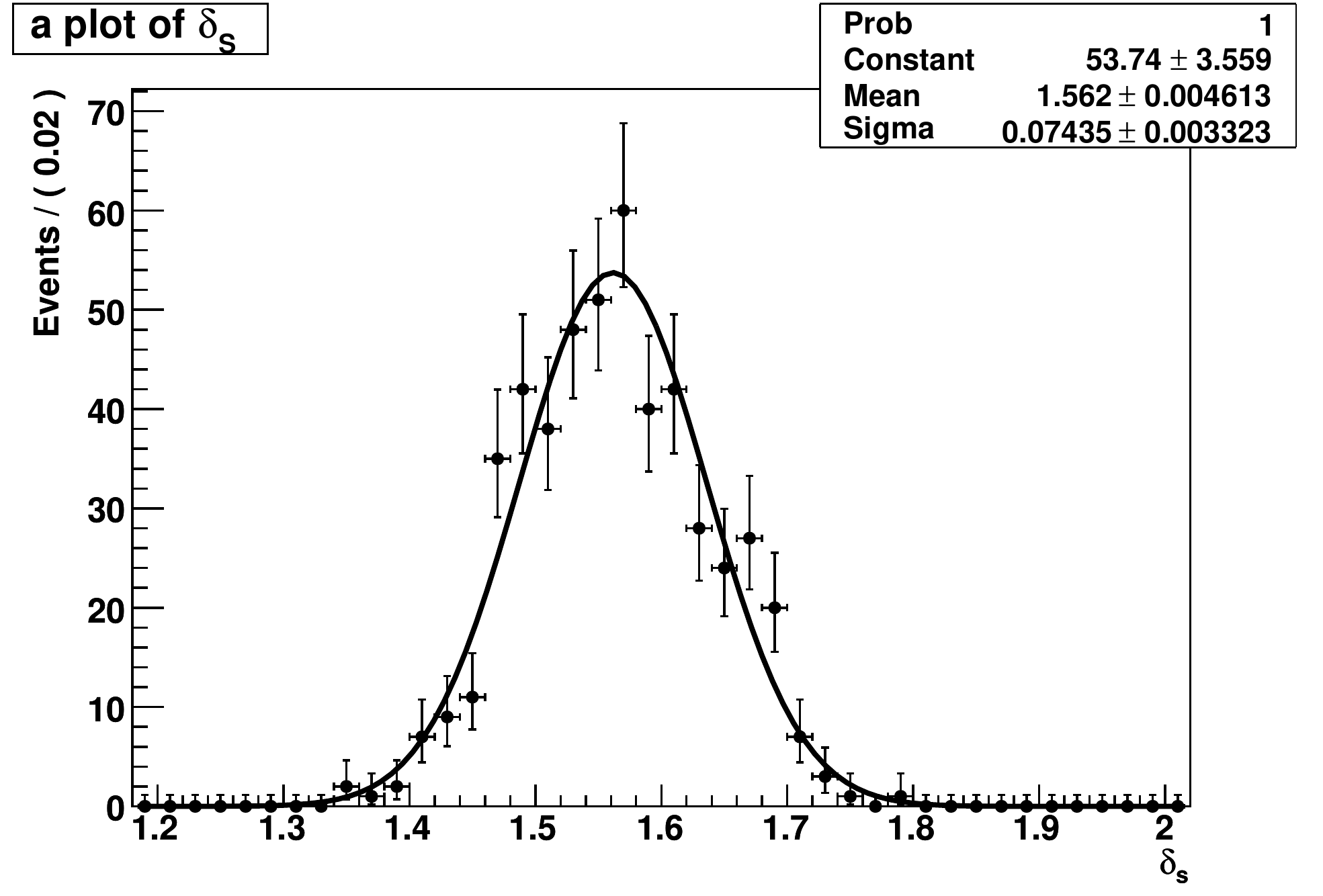}
\end{minipage}
\caption{ Distributions of the fitted values of $R_S$ and $\delta_S$ for the scenario $R_S=0.1, \delta_S=\pi/2, 2\betas =0.5$
without fixing $R_S$ to 0 in  fitting the data. }
\label{swAmpl}
\end{figure}

The results of these fits for the statistical error and mean value of the weak phase  $-2\betas$ 
are summarized in Table~\ref{tabRes0},~\ref{tabRes1} and~\ref{tabRes2} for several different 
scenarios with $-2\betas=-0.0368$,   $-2\betas=-0.2$ and $-2\betas=-0.5$, respectively.
As an example, in Figure~\ref{phisFit} we show the  distributions of the fitted values of $-2\betas$ for  $R_S=0.1, \delta_S=\pi/2$ and $-2\betas =-0.5$ for both  the S-wave fraction  $R_S$ fixed to zero and  $R_S$ left free in the fits.
In Figure~\ref{swAmpl} we show the distributions of the fitted values of $R_S$ and the strong phase of the S-wave 
component $\delta_S$ for the same case with $R_S$ left free in the fit.
It can be seen that when all parameters are fitted the results are unbiased,
but when it is wrongly assumed that $R_S = 0$, the result for $-2\beta_s$ acquires
a bias with regard to the true input value.

\begin{figure}[ht]
\begin{center}
   \includegraphics[angle=90,width=80mm]{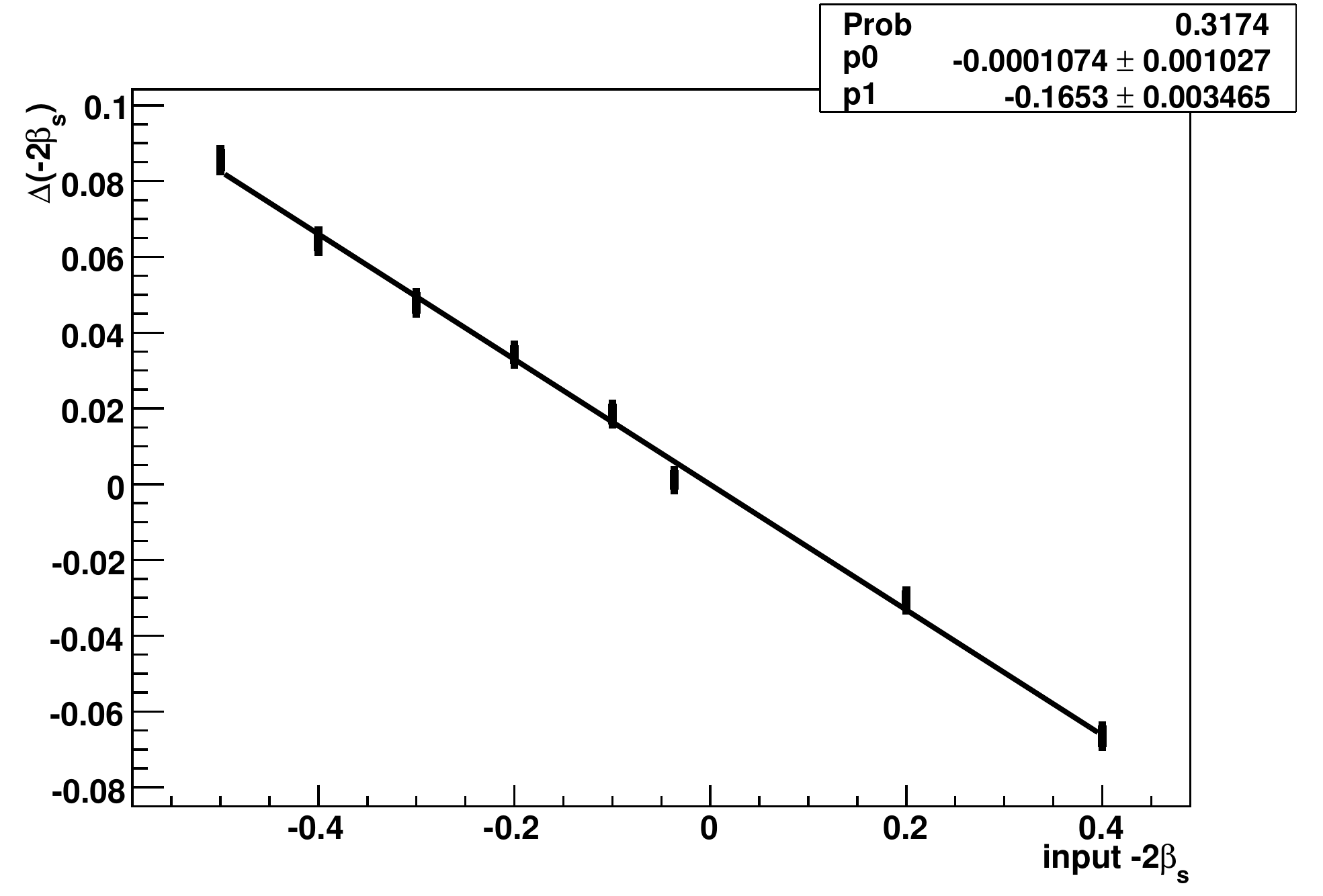}
\end{center}
\caption{The bias in $-2\betas$ from neglecting 
 an S-wave component with $R_S=0.1$ and  $\delta_S=\pi/2$
versus the value of $-2\betas$  used to generate the data sets. 
The bias is the difference of the mean of the fitted to the
generated $-2\betas$ values.
A linear fit is superimposed on the graph.
}
\label{scan}
\end{figure}

Figure~\ref{scan} shows the bias in $-2\betas$ from neglecting
 an S-wave component with $R_S=0.1$ and  $\delta_S=\pi/2$
versus the value of $-2\betas$  used to generate the data sets.
A linear dependence is observed, which demonstrates that the bias in $-2\betas$
is proportional to the true value of $-2\betas$.
 From Tables~\ref{tabRes0},~\ref{tabRes1} and~\ref{tabRes2} we observe biases for these scenarios which 
range from  $7-17\%$ in the measurement of  $2\betas$   
if an S-wave component is present, but left unaccounted for in the fits. The bias moves the measured  value of $2\betas$ towards zero. This implies
that the neglected  CP-odd S-wave contribution has a bigger probability to be mis-identified as the CP-even longitudinal or  
parallel components than as the CP-odd perpendicular component.  Therefore, although the bias from neglecting an S-wave contribution
 is unlikely to lead to false signal of new physics, it will cause a  loss of sensitivity to new physics. 
On the other hand, including the S-wave in the fit removes the bias
in the central value of  $2\betas$  at a cost of an increase of less than $15\%$ in the statistical error.

\begin{table}[h]
\footnotesize
\begin{center}
  \begin{tabular}{|l|c|c|}
  \hline
  & Float $R_S$ in fit & Fix $R_S$ to 0 in  fit  \tabularnewline
  \hline
  $R_S = 0$ &   & $\sigma(2\betas)= 0.045$, $\rm Mean(2\betas)= 0.038 $  \tabularnewline
  \hline
  $R_S = 0.1, \delta_S = \pi/2$ & $\sigma(2\betas)= 0.048$, $\rm Mean(2\betas)= 0.035 $  & 
                                  $\sigma(2\betas)= 0.045$, $\rm Mean(2\betas)= 0.032 $  \tabularnewline
  \hline
  $R_S = 0.1, \delta_S = 0$ & $\sigma(2\betas)= 0.054$, $\rm Mean(2\betas)= 0.040 $  & 
                              $\sigma(2\betas)= 0.048$, $\rm Mean(2\betas)= 0.036 $  \tabularnewline
  \hline
  $R_S = 0.05, \delta_S = \pi/2$ & $\sigma(2\betas)= 0.048$, $\rm Mean(2\betas)= 0.040 $  & 
                                   $\sigma(2\betas)= 0.045$, $\rm Mean(2\betas)= 0.036 $  \tabularnewline
  \hline
  $R_S = 0.05, \delta_S = 0$ & $\sigma(2\betas)= 0.055$, $\rm Mean(2\betas)= 0.038 $  & 
                               $\sigma(2\betas)= 0.047$, $\rm Mean(2\betas)= 0.032 $  
\tabularnewline
  \hline
  \end{tabular}
\end{center}
\normalsize
\caption{Statistical errors and mean values of $2\betas$ from 500 fits for different scenarios with  $2\betas = 0.0368$. 
 The errors on $\sigma(2\betas)$ and mean(2\betas) are approximately 0.003 and 0.002, respectively. 
 The same data sets are used to obtain the results in the second and third columns.
}
\label{tabRes0}
\end{table}

\begin{table}[h]
\footnotesize
\begin{center}
  \begin{tabular}{|l|c|c|}
  \hline
  & Float $R_S$ in fit & Fix $R_S$ to 0 in  fit  \tabularnewline
  \hline
  $R_S = 0$ &   & $\sigma(2\betas)= 0.044$, $\rm Mean(2\betas)= 0.198 $  \tabularnewline
  \hline
  $R_S = 0.1, \delta_S = \pi/2$ & $\sigma(2\betas)= 0.052$, $\rm Mean(2\betas)= 0.199 $  &
                                  $\sigma(2\betas)= 0.047$, $\rm Mean(2\betas)= 0.166 $  \tabularnewline
  \hline
  $R_S = 0.1, \delta_S = 0$ & $\sigma(2\betas)= 0.056$, $\rm Mean(2\betas)= 0.202 $  &
                              $\sigma(2\betas)= 0.049$, $\rm Mean(2\betas)= 0.170 $  \tabularnewline
  \hline   
  $R_S = 0.05, \delta_S = \pi/2$ & $\sigma(2\betas)= 0.049$, $\rm Mean(2\betas)= 0.197 $  &
                                   $\sigma(2\betas)= 0.048$, $\rm Mean(2\betas)= 0.182 $  \tabularnewline
  \hline   
  $R_S = 0.05, \delta_S = 0$ & $\sigma(2\betas)= 0.053$, $\rm Mean(2\betas)= 0.198 $  &
                               $\sigma(2\betas)= 0.048$, $\rm Mean(2\betas)= 0.180 $

\tabularnewline
  \hline
  \end{tabular}
\end{center}
\normalsize
\caption{Statistical errors and mean values of $2\betas$ from 500 fits for different scenarios with  $2\betas = 0.2$. 
 The errors on $\sigma(2\betas)$ and mean(2\betas) are approximately 0.003 and 0.002, respectively.
 The same data sets are used to obtain the results in the second and third columns.
}
\label{tabRes1}
\end{table}

\begin{table}[h]
\footnotesize
\begin{center}
  \begin{tabular}{|l|c|c|}
  \hline
  & Float $R_S$ in fit & Fix $R_S$ to 0 in  fit  \tabularnewline
  \hline
  $R_S = 0$ &   & $\sigma(2\betas)= 0.051$, $\rm Mean(2\betas)= 0.501 $  \tabularnewline
  \hline
  $R_S = 0.1, \delta_S = \pi/2$ & $\sigma(2\betas)= 0.059$, $\rm Mean(2\betas)= 0.501 $  &
                                  $\sigma(2\betas)= 0.053$, $\rm Mean(2\betas)= 0.415 $  \tabularnewline
  \hline
  $R_S = 0.1, \delta_S = 0$ & $\sigma(2\betas)= 0.061$, $\rm Mean(2\betas)= 0.501 $  &
                              $\sigma(2\betas)= 0.052$, $\rm Mean(2\betas)= 0.417 $  \tabularnewline
  \hline
  $R_S = 0.05, \delta_S = \pi/2$ & $\sigma(2\betas)= 0.051$, $\rm Mean(2\betas)= 0.506 $  &
                                   $\sigma(2\betas)= 0.048$, $\rm Mean(2\betas)= 0.463 $  \tabularnewline
  \hline
  $R_S = 0.05, \delta_S = 0$ & $\sigma(2\betas)= 0.053$, $\rm Mean(2\betas)= 0.501 $  &
                               $\sigma(2\betas)= 0.049$, $\rm Mean(2\betas)= 0.461 $
\tabularnewline
  \hline
  \end{tabular}
\end{center}
\normalsize
\caption{Statistical errors and mean values of $2\betas$ from 500 fits for different scenarios with  $2\betas = 0.5$. 
 The errors on $\sigma(2\betas)$ and mean(2\betas) are approximately 0.003 and 0.002, respectively.
 The same data sets are used to obtain the results in the second and third columns.
}
\label{tabRes2}
\end{table}

\cleardoublepage

\section{Measuring  $\cos2\betas$  }\label{cos2betas}
\setcounter{equation}{0}

In  Eq.~\ref{Eqbsrate} and~\ref{EqbsbarRate} one observes  that 
the differential decay rates are invariant under the transformation
\begin{equation}\label{EqAmbig}
\left( \delta_{||}-\delta_0, \delta_{\perp}-\delta_0, \delta_S-\delta_0, -2\betas, \DGs   \right)  \leftrightarrow
 \left( \delta_0-\delta_{||}, \pi+\delta_0-\delta_{\perp}, \delta_0-\delta_S, \pi -(- 2\betas), -\DGs \right)\,.
\end{equation}
As a consequence the measurement of $2\betas$ is subject to  a two-fold ambiguity,
which is equivalent to $\cos 2\betas$ transforming into  $-\cos 2\betas$. 
A measurement of $\cos 2\betas$  including its sign would allow us to resolve this ambiguity.

\begin{figure}[ht]
    \vfill\begin{minipage}{0.5\linewidth}
   \includegraphics[angle=90,width=80mm]{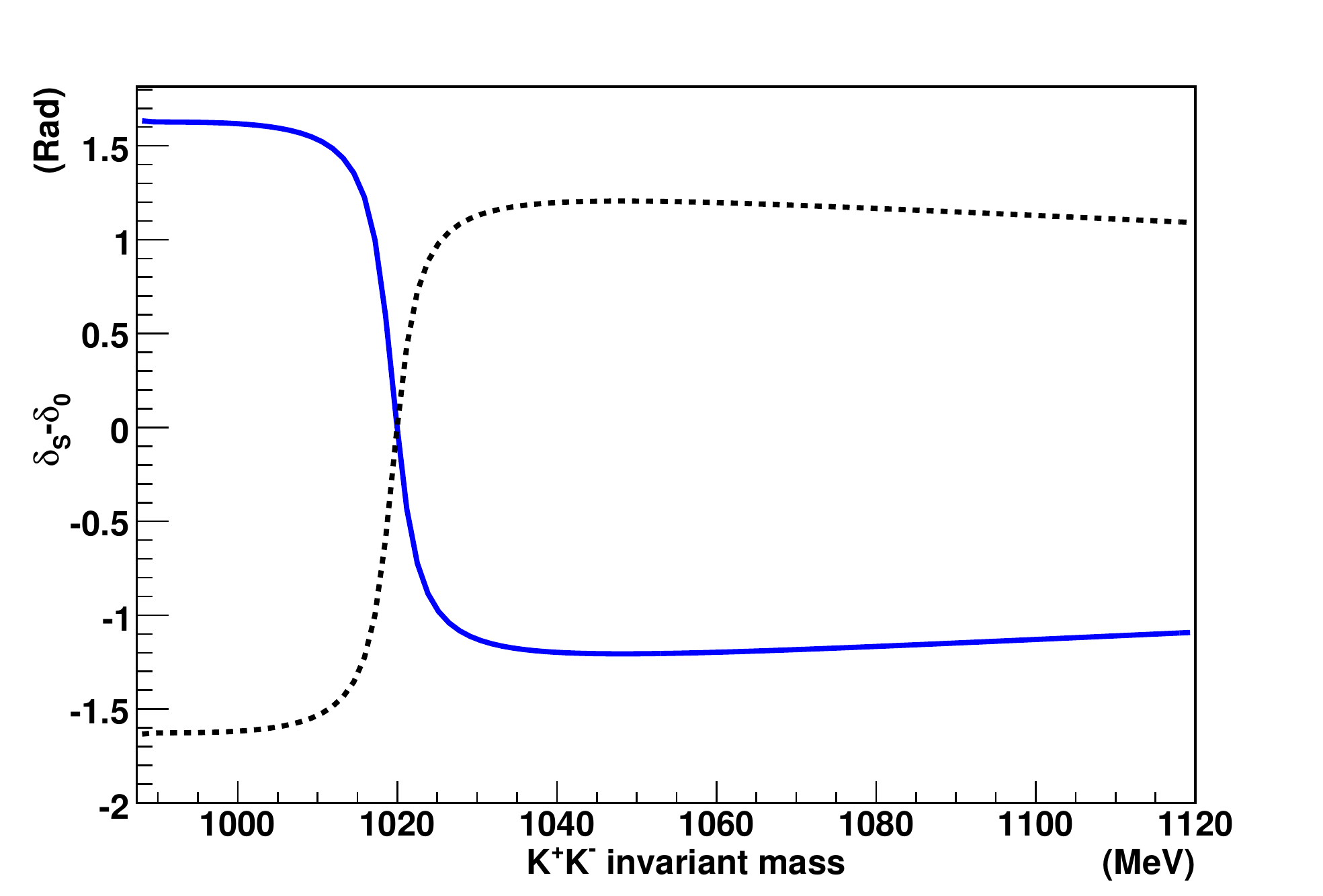}
\end{minipage} 
\begin{minipage}{0.5\linewidth}
   \includegraphics[angle=90,width=80mm]{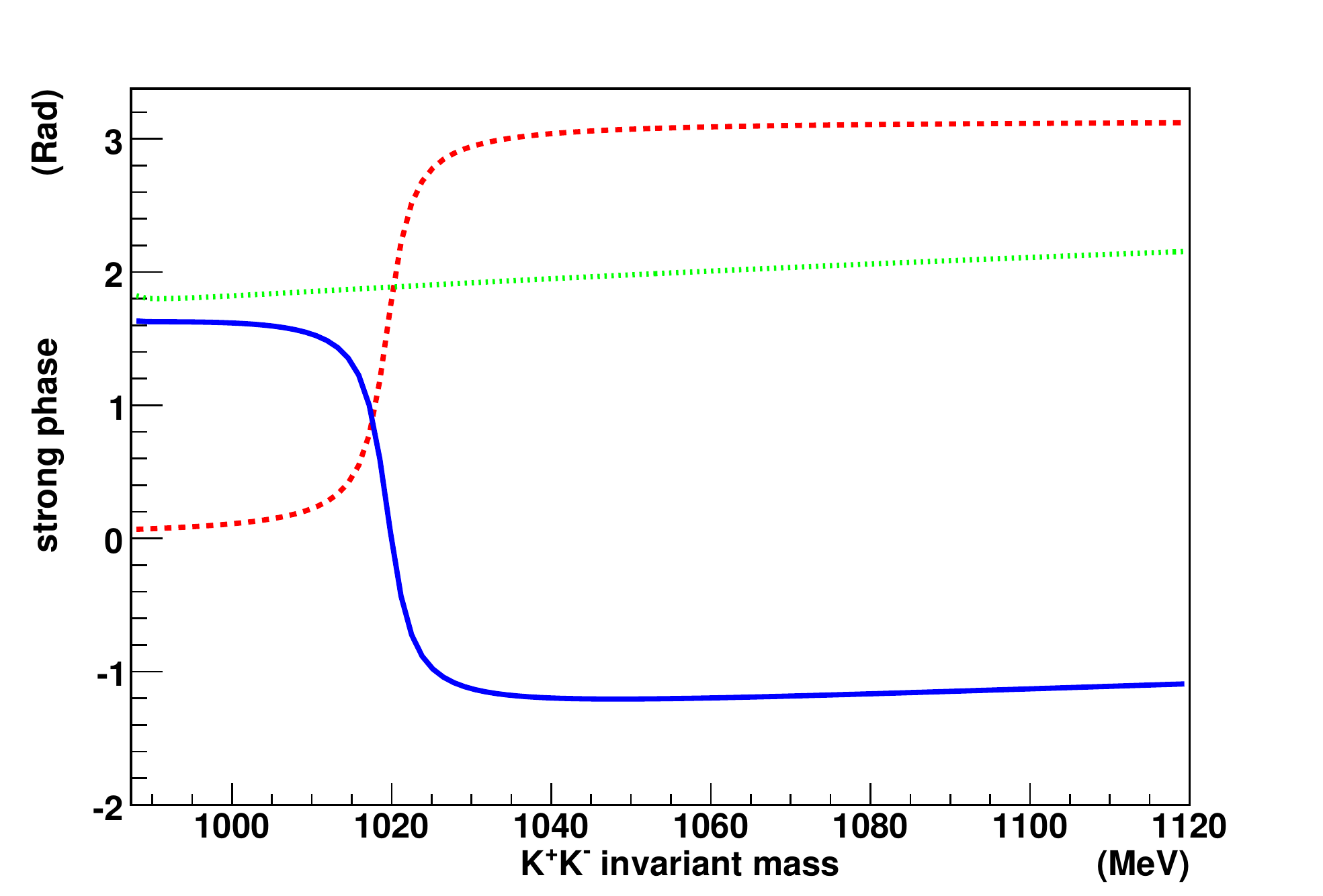}
\end{minipage}
\caption{An example to illustrate the  dependence of the strong phase of the  S-wave $\delta_S$,
 of the strong phase of the longitudinal P-wave $\delta_0$,  and of their difference  $\delta_S-\delta_0$, on the \KpKm\ mass.
Left: the solid blue curve is the physical solution for $\delta_S-\delta_0$ and the dashed black curve shows the mirror solution.
Right: the dashed red,  dotted green and  solid blue  curves are
 for $\delta_0$,  $\delta_S$, and  $\delta_S-\delta_0$, respectively.
 }
\label{phases}
\end{figure}

If the interference between the P-wave and S-wave amplitudes were to be significant
in the \particle{\phi (1020)} mass region,
we could use this effect to measure $\cos 2\betas$, in the same way as BaBar measured 
$\cos 2\beta$ in \decay{\Bd}{\Jpsi \KS\piz} \cite{BabarCos2beta}. 
This requires measuring $\delta_S -\delta_0$,
the strong phase difference between the S-wave
and the longitudinal P-wave, as a function of the \KpKm\ mass in the  \particle{\phi (1020)} mass region.
When plotting this function,  two branches are expected with 
each corresponding to a different solution for the weak phase (see  Figure~\ref{phases} left).
It is straightforward to choose the physical solution since the phase of the P-wave Breit-Wigner amplitude 
 is expected to rise  rapidly through the \particle{\phi (1020)} mass region (dashed red curve in  Figure~\ref{phases} right), 
while the phase of the S-wave amplitude, which can be described either by a coupled channel Breit-Wigner function
in case of  an \particle{f_0} contribution or by a constant term in case of a non-resonant contribution,
 is expected to vary relatively slowly 
(dotted green curve in  Figure~\ref{phases} right),
resulting in  $\delta_S - \delta_0$ rapidly falling with increasing \KpKm\ mass
(solid blue curves in  Figure~\ref{phases}).

\begin{figure}[ht]
\begin{center}
   \includegraphics[angle=90,width=80mm]{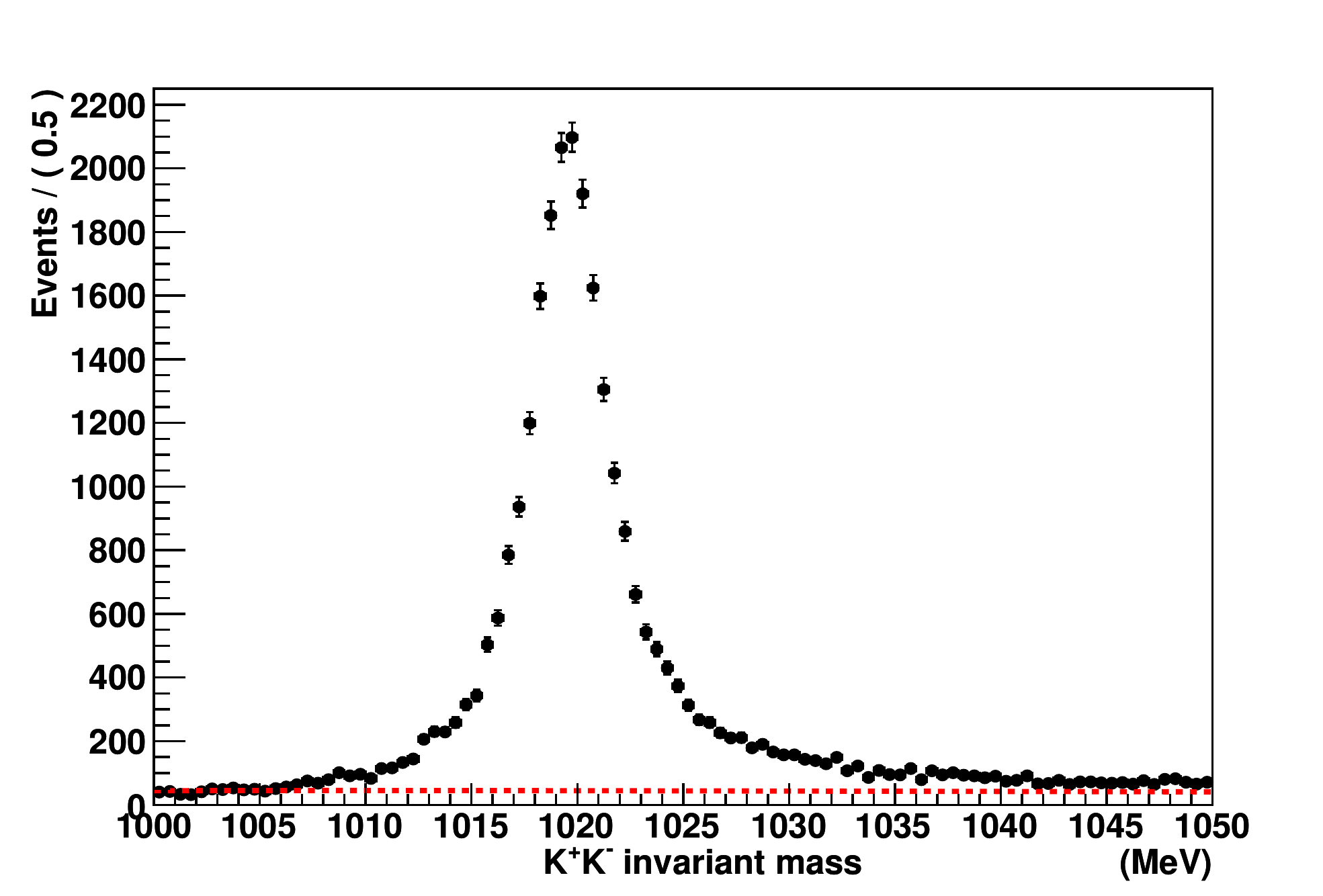}
\end{center}
\caption{The data points correspond to  the  \KpKm\ mass distribution of a generated sample of \decay{\Bs}{\Jpsi \KpKm} events
including $10\%$  \particle{f_0}  contribution in the mass region.  The dotted red curve indicates the  \particle{f_0}  contribution.
}
\label{kkmass}
\end{figure}

Below we use a Monte Carlo simulated toy data set
 to demonstrate the feasibility of this method in measuring the sign of $\cos2\betas$. 
We generate 30000 \decay{\Bs}{\Jpsi \KpKm} events in the  \KpKm\ mass  region between 1 and 
1.05 $\GeVcc$, roughly corresponding to 0.5\invfb\ of integrated luminosity. The P-wave  and  \particle{f_0} contributions are included coherently.
The values of the parameters used to generate the toy data set are the same as in Table~\ref{tabPara} 
except that we set  $-2\betas=-0.0368$, and that the values of both $R_S$ and $\delta_S$  depend on
the \KpKm\ mass. The  \particle{f_0}  contribution 
accounts for about $10\%$ of the total decay rate in the given mass region, as is shown in Figure~\ref{kkmass}.

\begin{figure}[ht]
\begin{center}
   \includegraphics[angle=90,width=80mm]{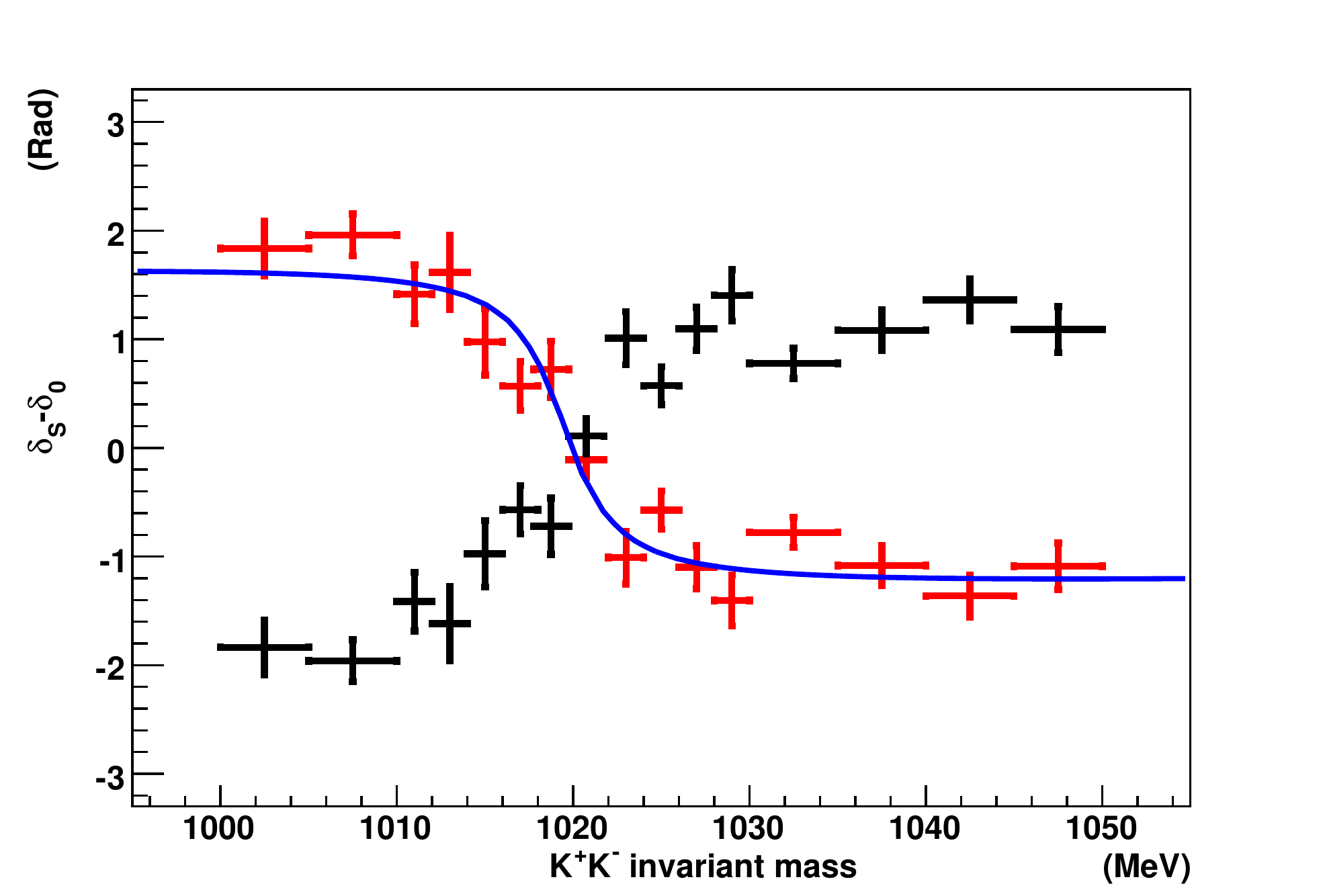}
\end{center}
\caption{The fitted values of $\delta_S-\delta_0$ versus  \KpKm\ mass are shown in red and black data points, corresponding to 
opposite values of $\cos2\betas$. The blue curve shows the dependence of $\delta_S-\delta_0$ on  \KpKm\ mass
implemented  in simulation.
}
\label{solutions}
\end{figure}

The data sample is divided into bins in the  \KpKm\ mass. For each bin $i$, two parameters $\delta_{S,i}$ and $R_{S,i}$ 
are used to represent the average strong phase and the fraction of the   \particle{f_0}  contribution. 
Both $\sin2\betas$  and $\cos2\betas$ are treated as independent free parameters. 
Common free parameters $\sin2\betas$, $\cos2\betas$, $R_{||}$, $R_{\perp}$, $\delta_{||}$, $\delta_{\perp}$, 
$\Gamma_s$ and $\Delta \Gamma_s$ are used for all bins.
Note that we still adopt the convention $\delta_0 =0 $ as only the relative phase differences  in each bin can be measured.
A combined fit to the time-dependent angular distributions of all the bins is performed to  extract these free parameters.
The fitted values of the strong phase difference 
$\delta_S-\delta_0$ versus the  \KpKm\ mass are plotted in Figure~\ref{solutions}. 
The two branches correspond to opposite values of $\cos2\betas$. Just as expected, the branch corresponding to the
true solution decreases rapidly around the nominal \particle{\phi (1020)}  mass. Choosing this branch leads to the unique
 solution

\begin{equation}\label{cos2betas}
 \sin2\betas = 0.043 \pm 0.05, \, \, \, \,
 \cos2\betas = 1.05 \pm 0.08 \,,
\end{equation}
which gives the ambiguity-free result
\begin{equation}\label{2betas}
 -2\betas = -0.043 \pm 0.05 \,.
\end{equation}
In this example, the measured $-2\betas$ is separated from $\pi-(-2\betas)$ by $13\sigma$,  therefore the discrete ambiguity in $2\betas$ is 
completely resolved. Although the actual measurement precision in $\cos2\betas$ will depend on the size of the \particle{f_0}  contribution
as well as background, the possibility to resolve the ambiguity in  $-2\betas$ using this method
is very promising.

\section{Conclusions}

In the decay \decay{\Bs}{\Jpsi \KpKm} we expect that a \KpKm\ S-wave contribution 
in the narrow \particle{\phi (1020)} mass region could be as large as $10\%$. 
The full differential decay rates for this decay including the S-wave contribution have been presented.
We have considered a range of scenarios which include S-wave components of $5$\% and $10$\%.
We have shown that within these scenarios,  if  an S-wave component is ignored in the analysis, 
the  measurement of the weak phase $-2\betas$  would be biased by between  $7\%$ and $17$\% towards zero.
We have demonstrated that by properly allowing for this  S-wave component in the fit,
an unbiased measurement of  $2\betas$  may be obtained with a slightly increased statistical
error.  Finally, we have shown that the interference between the \KpKm\ S-wave  and P-wave
amplitudes can be used to resolve the two-fold ambiguity in the measurement of the
weak phase  $-2\betas$.

\section*{Acknowledgments}

The authors would like to acknowledge  the LHCb colleagues for useful and stimulating discussions. We
particularly thank  Tim Gershon, Olivier Leroy and Guy Wilkinson for valuable suggestions.



\end{document}